\documentclass[11pt,german,english]{article}
\usepackage[T1]{fontenc}
\usepackage[latin9]{inputenc}
\usepackage[a4paper]{geometry}
\usepackage{color}
\usepackage{amsmath}
\usepackage{amssymb}
\usepackage{graphicx}

\makeatletter

\newcommand{\noun}[1]{\textsc{#1}}

\newcommand{\lyxaddress}[1]{
\par {\raggedright #1
\vspace{1.4em}
\noindent\par}
}

\usepackage{lipsum}

\let\OLDthebibliography\thebibliography
\renewcommand\thebibliography[1]{
  \OLDthebibliography{#1}
  \setlength{\parskip}{0pt}
  \setlength{\itemsep}{0pt plus 0.3ex}
}
\usepackage{amssymb}

\usepackage[bottom]{footmisc}

\makeatother

\usepackage{babel}
\begin{document}

\title{Generalizations of polylogarithms for Feynman integrals\date{}}

\author{Christian Bogner}

\maketitle

\lyxaddress{\begin{center}
\emph{Institut f\"ur Physik, Humboldt-Universit\"at zu Berlin,
}\\
\emph{D - 10099 Berlin, Germany}
\par\end{center}}
\begin{abstract}
In this talk, we discuss recent progress in the application of generalizations
of polylogarithms in the symbolic computation of multi-loop integrals.
We briefly review the Maple program MPL which supports a certain approach
for the computation of Feynman integrals in terms of multiple polylogarithms.
Furthermore we discuss elliptic generalizations of polylogarithms
which have shown to be useful in the computation of the massive two-loop
sunrise integral. 
\end{abstract}

\section{Motivation: Multiple polylogarithms and Feynman integrals}

Classical polylogarithms $\textrm{Li}_{n}$ are obtained as a generalization
of the logarithm function 
\[
\textrm{Li}_{1}(z)=-\ln(1-z)=\sum_{j=1}^{\infty}\frac{z^{j}}{j},\,\,\left|z\right|<1,
\]
by allowing for higher integer powers of the summation variable in
the denominator:
\[
\textrm{Li}_{n}(z)=\sum_{j=1}^{\infty}\frac{z^{j}}{j^{n}},\,\,\left|z\right|<1.
\]
These functions can be expressed in terms of integrals. For the dilogarithm,
Leibniz \cite{Lei} already found the identity 
\[
\textrm{Li}_{2}(z)=-\int_{0}^{z}\frac{dx}{x}\ln(1-x).
\]
\foreignlanguage{german}{In general, for weights $n\geq2,$ we have
\begin{eqnarray}
\textrm{Li}_{n}(z) & = & \int_{0}^{z}\frac{dx}{x}\textrm{Li}_{n-1}(x).\label{eq:Li_n integral}
\end{eqnarray}
If we write all integrations on the right-hand side of this equation
explicitly, we obtain an iterated integral
\begin{equation}
\textrm{Li}_{n}(z)=\int_{0}^{z}\frac{dx_{n}}{x_{n}}...\int_{0}^{x_{3}}\frac{dx_{2}}{x_{2}}\int_{0}^{x_{2}}\frac{dx_{1}}{1-x_{1}}.\label{eq:class poly interated int}
\end{equation}
In this talk, we denote iterated integrals by 
\[
\left[\omega_{r}|...|\omega_{2}|\omega_{1}\right]=\int_{0}^{z}\omega_{r}(x_{r})...\int_{0}^{x_{3}}\omega_{2}(x_{2})\int_{0}^{x_{2}}\omega_{1}(x_{1})
\]
where the $\omega_{i}$ are differential 1-forms in some given set.
In eq. \ref{eq:class poly interated int} we see that the set 
\[
\Omega_{\textrm{P}}=\left\{ \frac{dx}{x},\,\frac{dx}{1-x}\right\} 
\]
suffices to construct the classical polylogarithms.}

\selectlanguage{german}%
Generalizations of polylogarithms can be obtained by either generalizing
the terms in the sum representation or by extending the set of differential
1-forms. In both ways, one arrives at multiple polylogarithms. They
are defined as the series \cite{Gon2,Gon1}
\[
\textrm{Li}_{n_{1},...n_{k}}\left(z_{1},...,z_{k}\right)=\sum_{0<j_{1}<...<j_{k}}\frac{z_{1}^{j_{1}}...z_{k}^{j_{k}}}{j_{1}^{n_{1}}...j_{k}^{n_{k}}}\textrm{ for }|z_{i}|<1
\]
and they can be expressed  in terms of iterated integrals known as
hyperlogarithms \cite{Poi,Lap1,Lap2}. These are obtained from differential
1-forms of the set 
\begin{equation}
\Omega_{\textrm{Hyp}}=\left\{ \left.\frac{dx}{x},\,\frac{dx}{x-y_{i}}\right|i=1,...,k\right\} .\label{eq:1-forms Hyperlog}
\end{equation}

Some powerful methods and computer programs for the analytical computation
of Feynman integrals rely on either the sum representation or on an
integral representation of multiple polylogarithms. In section \ref{sec:Iterated-integrals-and}
we review the computer program MPL which supports an approach based
on iterated integrals. For Feynman integrals which can not be expressed
in terms of multiple polylogarithms, we are in search of alternatives.
In section \ref{sec:Elliptic-generalizations} we briefly recall the
concept of elliptic functions and in section \ref{sec:The-massive-sunrise}
we discuss an elliptic generalization of polylogarithms which arises
from the computation of the massive sunrise integral.

\selectlanguage{english}%

\section{Iterated integrals and the program MPL\label{sec:Iterated-integrals-and}}

As an alternative to hyperlogarithms, we consider a class of iterated
integrals over differential 1-forms in the set 
\begin{equation}
\Omega_{\textrm{MPL}}=\left\{ \left.\frac{dx_{1}}{x_{1}},...,\frac{dx_{k}}{x_{k}},\,\frac{d\left(p_{a,b}\right)}{p_{a,b}}\right|1\leq a\leq b\leq k\right\} \label{eq:1-forms MPL}
\end{equation}
where 
\begin{equation}
p_{a,b}=\prod_{a\leq i\leq b}x_{i}-1.\label{eq:polynomials p_a,b}
\end{equation}
\foreignlanguage{german}{In order to obtain a framework of well-defined
functions of the $k$ variables $x_{1},\,...,\, x_{k},$ we construct
only iterated integrals which are homotopy invariant. In general,
an iterated integral admits this property, if and only if it satisfies
the condition \cite{Che}
\begin{equation}
\mathcal{D}\left[\omega_{1}|...|\omega_{m}\right]=0\label{eq:int condition}
\end{equation}
where the operator $\mathcal{D}$ is defined by 
\begin{equation}
\mathcal{D}\left[\omega_{1}|...|\omega_{m}\right]=\sum_{i=1}^{m}[\omega_{1}|...|\omega_{i-1}|d\omega_{i}|\omega_{i+1}|...\omega_{m}]+\sum_{i=1}^{m-1}[\omega_{1}|...|\omega_{i-1}|\omega_{i}\wedge\omega_{i+1}|...|\omega_{m}].\label{eq:integrability}
\end{equation}
}For example, among the two integrals 
\[
I_{1}=\left[\left.\frac{dx_{3}}{x_{3}}+\frac{dx_{2}}{x_{2}}\right|\frac{d\left(x_{2}x_{3}\right)}{x_{2}x_{3}-1}\right],\, I_{2}=\left[\left.\frac{dx_{3}}{x_{3}}\right|\frac{d\left(x_{2}x_{3}\right)}{x_{2}x_{3}-1}\right],
\]
only $I_{1}$ is homotopy invariant while $I_{2}$ fails eq. \ref{eq:int condition}.
For any number of variables $k$, we can apply algorithms described
in \cite{BogBro1,BogBro2} to construct a basis of all homotopy invariant
iterated integrals over 1-forms in $\Omega_{\textrm{MPL}}.$ Together
with certain boundary conditions at a tangential basepoint (see \cite{BogBro2}),
this construction provides a $\mathbb{Q}$-vectorspace $V\left(\Omega_{\textrm{MPL}}\right)$
of functions, including the class of multiple polylogarithms.

MPL \cite{Bog} is a Maple program for computations with this class
of functions. Its main algorithms \cite{BogBro2} rely on the mathematical
theory developed in \cite{Bro1}. One of the main purposes of the
program is the computation of definite integrals of the type \foreignlanguage{german}{
\begin{equation}
I=\int_{0}^{1}dx_{n}\frac{q}{\prod_{j}p_{j}^{a_{j}}}f\label{eq:integral cubical type}
\end{equation}
where $f\in V\left(\Omega_{\textrm{MPL}}\right),$ $q$ is some arbitrary
polynomial in $x_{n}$, all $a_{j}\in\mathbb{N}$ and all $p_{j}$
are polynomials of the type of eq. \ref{eq:polynomials p_a,b}. For
example, the program computes analytically 
\[
\int_{0}^{1}dx_{1}\int_{0}^{1}dx_{2}\int_{0}^{1}dx_{3}\frac{x_{1}^{4}\left(1-x_{1}\right)^{4}x_{2}^{9}\left(1-x_{2}\right)^{4}x_{3}^{4}\left(1-x_{3}\right)^{4}}{\left(1-x_{1}x_{2}\right)^{5}\left(1-x_{2}x_{3}\right)^{5}}=-\frac{11424695}{144}+66002\zeta(3).
\]
Such integrals appear in various contexts. Examples are given in \cite{Bro4,Bog,BogBro2}.}

\selectlanguage{german}%
The other main purpose of the program MPL is \textcolor{black}{the
analytical computation of a certain class of scalar Feynman integrals.
For some Feynman graph $G,$ consider the $D$-dimensional, scalar
$L$-loop integral }
\begin{equation}
I(\Lambda)=\Gamma\left(\nu-LD/2\right)\left(\prod_{i=1}^{N}\int_{0}^{\infty}\frac{dx_{i}x_{i}^{\nu_{i}-1}}{\Gamma(\nu_{i})}\right)\delta\left(H\right)\frac{\mathcal{{\color{red}{\color{black}U}}}^{\nu-(L+1)D/2}}{\left(\mathcal{{\color{red}{\color{black}F}}}\left(\Lambda\right)\right)^{\nu-LD/2}},\label{eq:Feynman integral}
\end{equation}
where $N$ is the number of edges of $G$, $\nu_{i}$ are integer
powers of the Feynman propagators, $\nu$ is the sum of all $\nu_{i},$
$\Lambda$ is a set of kinematical invariants and masses and $H=1-\sum_{i\in S}x_{i}$
for some choice of $S\subseteq\{1,...,N\}.$ The terms $\mathcal{U}$
and $\mathcal{F}$ are the Symanzik polynomials in the Feynman parameters
$x_{1},...,x_{N}$ (see e.g. \cite{BogWei2}). Applying the methods
of \cite{Pan3,ManPanSch,BroKre} we can expand such a possibly divergent
integral as a Laurent series in a parameter $\epsilon$ of dimensional
regularisation, 
\[
I=\sum_{j=-2L}^{\infty}I_{j}\epsilon^{j},
\]
such that the integrals $I_{j}$ are finite. The integrands of these
$I_{j}$ will involve Symanzik polynomials of $G$ and of related
graphs. 

In general, Symanzik polynomials are more complicated than the polynomials
of eq. \ref{eq:polynomials p_a,b}. Therefore, the computation involves
some additional steps. Before each integration, MPL attempts to express
the integrand in the form of eq. \ref{eq:integral cubical type} by
an appropriate change of variables. Then the integral is computed
and the result is mapped back to Feynman parameters, as a preparation
of the next integration. In order for all Feynman parameters to be
integrated out in this way, the (Symanzik) polynomials in the original
integrand have to satisfy the condition of linear reducibility as
discussed in \cite{Bro5,Bro6,Pan1}. This and two further conditions
can be checked by the program. If they are satisfied, the integral
can be computed automatically with MPL. Examples are given in \foreignlanguage{english}{\cite{Bog}
and in a manual obtained with the program.}

A similar approach is followed by Panzer's program HyperInt \cite{Pan2},
based on hyperlogarithms, which is publicly available as well, and
related methods are applied by programs discussed in \cite{Abl3,Abl4,Abl7,Abl8,Abletal,Broeetal}. 

What if a given Feynman integral does not satisfy the criterion of
linear reducibility? In some cases, this problem is just an artefact
of the parametrization and after some clever change of variables,
the above approach can still be applied%
\footnote{An example for such a case is the graph found to be irreducible in
\cite{BogLue} and later computed in \cite{HennSmi,Pan3}. %
}. However, there are as well Feynman integrals, which can not be expressed
in terms of multiple polylogarithms, no matter which parameters or
classes of iterated integrals we try to apply. For such Feynman integrals,
we have to turn to other frameworks of functions. The given success
with multiple polylogarithms suggests to give further generalizations
of polylogarithms a try.

\selectlanguage{english}%

\section{Elliptic generalizations\label{sec:Elliptic-generalizations}}

Let us recall the basic concept of an elliptic function.
\begin{figure}
\selectlanguage{german}%
\begin{centering}
\includegraphics[bb=0bp 420bp 612bp 742bp,clip,scale=0.3]{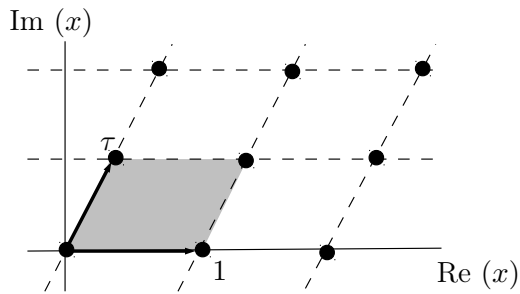}\put(-24, 5) {{ Re ${(x)} $}}%
\put(-184, 100) {{ Im ${(x)} $}}%
\put(-108, 6) {{ ${1} $}}%
\put(-150, 55) {{ ${\tau} $}}%
\par\end{centering}

\selectlanguage{english}%
\caption{A lattice in the complex plane\label{fig:lattice}}

\end{figure}
\foreignlanguage{german}{In the complex plane of a variable $x\in\mathbb{C}$
we consider the lattice $L=\mathbb{Z}+\tau\mathbb{Z}$, where $\tau\in\mathbb{C}$
with Im$(\tau)>0$ (the points in fig. \ref{fig:lattice}). A function
$f(x)$ is called elliptic with respect to $L$ if 
\begin{equation}
f(x)=f(x+\lambda)\textrm{ for }\lambda\in L.\label{eq:ellipticity}
\end{equation}
It makes sense to consider such a function $f$ only in one cell of
the lattice (the grey area in fig. \ref{fig:lattice}), as its behaviour
in all other cells are just copies. If $\tau$ is the quotient of
two periods $\psi_{1},\,\psi_{2}$ of an elliptic curve $E,$ this
cell of the lattice is isomorphic to $E$ and we can consider $f$
as a function on the elliptic curve. }

\selectlanguage{german}%
Now we introduce a change of variables, considering the function $f'(z)$
of $z\in\mathbb{C}^{\star}$ given by 
\[
f'\left(e^{2\pi ix}\right)=f\left(x\right).
\]
Clearly, if $f$ is elliptic with respect to $L,$ then with respect
to the new variable $z,$ eq. \ref{eq:ellipticity} implies 
\begin{equation}
f'\left(z\right)=f'\left(z\cdot q\right)\textrm{ where }q=e^{2\pi i\lambda}\textrm{ for }\lambda\in L.\label{eq:ellipticity z}
\end{equation}

\selectlanguage{english}%
Now there is a simple idea for the construction of such elliptic functions.
If $f'(z)$ can be defined with the help of some other function $g(z)$
as
\[
f'(z)=\sum_{n\in\mathbb{Z}}g\left(z\cdot q^{n}\right),
\]
it satisfies eq. \ref{eq:ellipticity z} by construction.

This concept can be applied to define elliptic generalizations of
polylogarithms. A first version of an elliptic dilogarithm was defined
in \cite{Blo} for the single-valued Bloch-Wigner dilogarithm. Later
the concept was generalized in various directions (see \cite{Zag,BeiLev,Lev,GanZag}).
Let us refer particularly to \cite{BroLev} where elliptic polylogarithms
of the form 
\begin{equation}
E_{m}(z)=\sum_{n\in\mathbb{Z}}u^{n}\textrm{Li}_{m}\left(z\cdot q^{n}\right)\label{eq:E Brown Levin}
\end{equation}
with a damping factor $u$ are considered and where the concept is
furthermore generalized to establish multiple elliptic polylogarithms. 

In the following section, a related class of functions appears in
the context of a Feynman integral.

\section{The massive sunrise integral\label{sec:The-massive-sunrise}}

The massive sunrise integral 
\[
S(D,\, t)=\int\frac{d^{D}k_{1}d^{D}k_{2}}{\left(i\pi^{D/2}\right)^{2}}\frac{1}{\left(-k_{1}^{2}+m_{1}^{2}\right)\left(-k_{2}^{2}+m_{2}^{2}\right)\left(-\left(p-k_{1}-k_{2}\right)^{2}+m_{3}^{2}\right)}
\]
is a Feynman integral which can not be expressed in terms of multiple
polylogarithms. This integral was extensively considered in the literature
\cite{Baileyetal,Bauetal,Bauetal2,Bauetal3,Beretal,BroFleTar,CafCzyRem,Caffoetal,CafGunRem,GroKoePiv,GroKoePiv2,LapRem,PozRem,RemTan,UssDav,RemTan2,Broadhu,KalKni,DavDel,DavSmi}.
In a recent computation of the case $D=2$ and equal masses, $m_{1}=m_{2}=m_{3},$
for the first time an elliptic polylogarithm was applied explicitly
to express a Feynman integral \cite{BloVan}. Here we discuss further
cases of the sunrise integral where elliptic generalizations of polylogarithms
arise.

At first, let us consider the integral with three different particle
masses as Laurent series at two and around four dimensions: 
\begin{eqnarray}
S(2-2\epsilon,\, t) & = & {\color{black}{\color{red}{\color{black}S^{(0)}(2,\, t)}}+{\color{black}{\color{red}{\color{blue}{\color{black}S^{(1)}(2,\, t)}}}}\epsilon}+\mathcal{O}\left(\epsilon^{2}\right),\label{eq:S(2)}\\
S(4-2\epsilon,\, t) & = & S^{(-2)}(4,\, t)\epsilon^{-2}+S^{(-1)}(4,\, t)\epsilon^{-1}+{\color{red}{\color{blue}{\color{black}S^{(0)}(4,\, t)}}}+\mathcal{O}(\epsilon).\label{eq:S(4)}
\end{eqnarray}
Here we have used $t=p^{2}.$ We begin with the result for exactly
$D=2$ dimensions, $S^{(0)}(2,\, t).$ In this case, the Feynman parametric
representation (eq. \ref{eq:Feynman integral}) of the sunrise integral
only involves the second Symanzik polynomial 
\[
\mathcal{F}=-x_{1}x_{2}x_{3}t+\left(x_{1}m_{1}^{2}+x_{2}m_{2}^{2}+x_{3}m_{3}^{2}\right)\left(x_{1}x_{2}+x_{2}x_{3}+x_{1}x_{3}\right)
\]
whose zero-set intersects the integration domain at three points $P_{1},\, P_{2},\, P_{3}.$
Together with each possible choice of one of these points as the origin,
this zero-set defines an elliptic curve. 

In \cite{AdaBogWei2} the following functions are introduced: 

\selectlanguage{german}%
\textcolor{black}{
\begin{equation}
\textrm{ELi}_{n;m}(x;y;q)=\sum_{j=1}^{\infty}\sum_{k=1}^{\infty}\frac{x^{j}}{j^{n}}\frac{y^{k}}{k^{m}}q^{jk}=\sum_{k=1}^{\infty}\frac{y^{k}}{k^{m}}\textrm{Li}_{n}(q^{k}x),\label{eq:ELi}
\end{equation}
}

\textcolor{black}{
\[
\textrm{E}_{n;m}(x;\, y;\, q)=
\]
\begin{equation}
\begin{cases}
\frac{1}{i}\left(\frac{1}{2}\textrm{Li}_{n}(x)-\frac{1}{2}\textrm{Li}_{n}(x^{-1})+\textrm{ELi}_{n;m}(x;y;q)-\textrm{ELi}_{n;m}(x^{-1};y^{-1};q)\right) & \textrm{ for }n+m\textrm{ even,}\\
\frac{1}{2}\textrm{Li}_{n}(x)+\frac{1}{2}\textrm{Li}_{n}(x^{-1})+\textrm{ELi}_{n;m}(x;y;q)+\textrm{ELi}_{n;m}(x^{-1};y^{-1};q) & \textrm{ for }n+m\textrm{ odd.}
\end{cases}\label{eq:E}
\end{equation}
Notice that these definitions are related to the basic ideas recalled
in section \ref{sec:Elliptic-generalizations} but slightly differ}%
\footnote{\textcolor{black}{For $n=2,$ the relation between these definitions
is made more explicit in \cite{AdaBogWei4}.}%
}\textcolor{black}{{} from the functions of eq. \ref{eq:E Brown Levin}.
By use of the differential equation of second order \cite{MueWeiZay1}
for $S^{(0)}(2,\, t)$, we obtain 
\begin{equation}
{\color{black}{\color{red}{\color{black}S^{(0)}\left(2,\, t\right)=\frac{\psi_{1}(q)}{\pi}\sum_{i=1}^{3}\textrm{E}_{2;\,0}(w_{i}(q);\,-1;\,-q)}}}\textrm{ where }q=e^{i\pi\frac{\psi_{2}}{\psi_{1}}}.\label{eq:S 0}
\end{equation}
Here $\psi_{1}$ and $\psi_{2}$ are periods of the elliptic curve
defined by $\mathcal{F},$ which are given by complete elliptic integrals
of the first kind \cite{AdaBogWei1}. The three arguments $w_{i}(q),$
$i=1,2,3,$ are directly related to the three intersection points
}\foreignlanguage{english}{$P_{1},\, P_{2},\, P_{3}$ (see \cite{AdaBogWei2}).}

\selectlanguage{english}%
While all terms in eq. \ref{eq:S 0} can be nicely related to the
underlying elliptic curve, the situation becomes considerably more
complicated for the higher coefficients of eqs. \ref{eq:S(2)} and
\ref{eq:S(4)}. Here the integrands under consideration depend on
both Symanzik polynomials. However, the functions defined in eqs.
\ref{eq:ELi} and \ref{eq:E} remain to be useful. Generalizing this
concept, we furthermore introduce the multi-variable functions \cite{AdaBogWei5}
\[
\textrm{ELi}_{n_{1},...,n_{l};m_{1},...,m_{l};2o_{1},...,2o_{l-1}}\left(x_{1},...,x_{l};y_{1},...,y_{l};q\right)
\]
\begin{equation}
=\sum_{j_{1}=1}^{\infty}...\sum_{j_{l}=1}^{\infty}\sum_{k_{1}=1}^{\infty}...\sum_{k_{l}=1}^{\infty}\frac{x_{1}^{j_{1}}}{j_{1}^{n_{1}}}...\frac{x_{l}^{j_{l}}}{j_{l}^{n_{l}}}\frac{y_{1}^{k_{1}}}{k_{1}^{m_{1}}}..\frac{y_{l}^{k_{l}}}{k_{l}^{m_{l}}}\frac{q^{j_{1}k_{1}+...+j_{l}k_{l}}}{\prod_{i=1}^{l-1}\left(j_{i}k_{i}+...+j_{l}k_{l}\right)^{o_{i}}}.\label{eq:E multi}
\end{equation}
By use of this set-up of functions, the coefficients $S^{(1)}(2,\, t)$
and ${\color{blue}{\color{black}S^{(0)}(4,\, t)}}$ are computed in
\cite{AdaBogWei3}. Furthermore it is shown in \cite{AdaBogWei5},
that in the case of equal masses, every higher coefficient of the
two-dimensional case $S(2-2\epsilon,\, t)$ can be expressed in terms
of these functions as well. This result includes an explicit algorithm
for the computation of these coefficients.

We want to point out that other elliptic generalizations of polylogarithms
recently found further applications to the two- and the three-loop
sunrise graph \cite{BloKerVan,BloKerVan2} and to integrals arising
in string theory \cite{Hokeretal,BroMafMatSch,BroMatSch}.

\section{Conclusions\label{sec:Conclusions}}

With their double nature as nested sums and iterated integrals, multiple
polylogarithms provide a very useful framework for the computation
of Feynman integrals. The Maple program MPL serves for the computation
of a certain class of Feynman integrals in terms of these functions.
The program is publicly available and supports computations with a
class of iterated integrals, which arise in other contexts as well. 

The massive two-loop sunrise integral is a Feynman integral which
can not be expressed in terms of multiple polylogarithms. For the
computation of various cases of this integral, a class of elliptic
generalizations of polylogarithms has shown to be useful. These and
other appearances of elliptic generalizations give rise to the hope,
that when we have to leave the realm of multiple polylogarithms, we
might not have to dispense with all of its advantages.

\end{document}